# The Data Dilemma: Authors' Intentions and Recognition of Research Data in Educational Technology Research

Sandra Schulz[1], Natalie Kiesler[2]

[1]University of Hamburg, Germany
[2]Nuremberg Tech, Germany

**Abstract**

Educational Technology (EdTec) research is conducted by multiple disciplines, some of which annually meet at the DELFI conference. Due to the heterogeneity of involved researchers and communities, it is our goal to identify categories of research data overseen in the context of EdTec research. Therefore, we analyze the author's perspective provided via EasyChair where authors specified whether they had research data to share. We compared this information with an analysis of the submitted articles and the contained research data. We found that not all research data was recognized as such by the authors, especially software and qualitative data, indicating a prevailing lack of awareness, and other potential barriers. In addition, we analyze the 2024 DELFI proceedings to learn what kind of data was subject to research, and where it is published. This work has implications for training future generations of EdTec researchers. It further stresses the need for guidelines and recognition of research data publications (particularly software, and qualitative data).

## 1 Introduction

*Do I, or do I not have research data to share?* – Answering this question may not be as easy as it sounds. In the context of the DELFI 2023 conference submission process, authors were requested to specify if they have any research data to share or if they intend to share them in the future. For 42% of the papers, including full research and demo papers, authors indicated not to have any data for publication (Kiesler, Röpke, et al. 2024). Yet, this information may be questionable, as empirical research and tools naturally produce respective data worth publishing. It was no surprise when, in 2024, Striewe asked: *What are we talking about when we talk about research data in Educational Technology (EdTec) research, what is relevant and when?* (Striewe 2024)

According to Kindling and Schirmbacher, "*By digital research data we mean [...] all digitally available data, that are generated during the research process or are its result*" (Kindling and Schirmbacher 2013). The definition of research data by the German Research Foundation (DFG), is more concrete, and includes, for example, "*measurement data, laboratory values, audiovisual information, texts, survey or observation data, methodological test procedures, and questionnaires. Compilations and simulations can likewise constitute a key outcome of academic research (. . . ). The same applies if the software is required to create or process research data.*" (German Research Foundation 2023) Thus, in the context of EdTec, research data may come in the form of source code, software artefacts; databases; knowledge graphs; text, audio, video and image data (Biernacka and Schulz 2022).

Despite these definitions, and the DELFI community's focus on original, empirical research and tools, there does not seem to be a consensus or awareness regarding research data, what it is, and publishing it. For this reason, it is our **goal** (1) to investigate DELFI 2024 authors' recognition of their research data as such, and (2) to identify published research data (what and where) in the context of the DELFI 2024 proceedings. The **contribution** of this work is an analysis of the DELFI community's recognition of research data and respective data publications. Thereby, we raise awareness and continue the discussion of Open Data practices. In the long run, this will help our community save resources, and value data as the hidden treasure that it is (European Commission and Directorate-General for Research and Innovation 2018).

## 2 Related Work

Open Science and Open Data principles have not yet gained traction within the DELFI community. It is, therefore, no surprise that only a handful of studies and papers are dedicated to this gap in research and practice. For example, Melkamu Jate and Striewe (2023) investigated the DELFI proceedings between 2018 and 2022. They analyzed the extent to which the publications make research data findable and accessible in alignment with the FAIR principles (Wilkinson et al. 2016). Interestingly, Melkamu Jate and Striewe (2023) did not denote software as research data and analyzed it separately. Their results showed that 51 to 68 % of the 2018–2022 publications contained research data (other than software), and 9 to 32 % of them were accessible (Melkamu Jate and Striewe 2023). Software artifacts were evident in a similar frequency; ranging between 58 % and 70 %. However, software was more accessible than the other research data (ranging between 23 % and 45 % over the years) (Melkamu Jate and Striewe 2023). The study also revealed that authors rarely used dedicated services, such as Zenodo or OSF. Moreover, software as the core of demo and tools papers was hardly published and accessible (Melkamu Jate and Striewe 2023).

A follow-up study analyzed the DELFI 2023 authors' intentions to publish their research data. It showed that for 28 out of the 66 accepted papers, authors claimed to have no research data (Kiesler, Röpke, et al. 2024). For 19 out of the 66 papers, authors had no intention to publish their data. Only 3 author teams had published their research data at the time they submitted to the conference, and respective repositories were available. Yet, none of the full research papers were among those who had published their research data upon submission. For 16 out of the 66 papers, authors had the intention to publish their data when they submitted to EasyChair, but only 4 of the 16 followed-up, and uploaded code repositories to GitLab and GitHub (Kiesler, Röpke, et al. 2024). Kiesler, Röpke, et al. (2024) thus showed the absence of research data during submission, the review process, and upon publication. Some DELFI 2023 authors seemed to lack awareness of their data. For example, 9 of the 21 demo author teams claimed to have no data at all (Kiesler, Röpke, et al. 2024) – although demos typically involve prototypes.

All of these findings are in line with a discussion led by Kiesler and Schiffner (2022) who outlined the lack of recognition of software as research data, criticizing structural issues. Among them is a lack of strategies and guidelines for the publication of research data, respective infrastructure and formats; publication categories at conferences; and incentives for authors to share and reuse data (Kiesler and Schiffner 2022). Schneider, Limbu, and Kiesler (2025) go even further, discussing the lack of mature research data and prototypes as one of the root causes hindering the impact of EdTec research. In an attempt to clarify the concept of research data in the EdTec community, Striewe (2024) presents a comprehensive overview of data artifacts gathered and generated in the context of EdTec research. It aligns with the DFG's definition of research data (German Research

Foundation 2023), and incorporates involved actors, processes, systems, and their interactions as potential sources of research data (Striewe 2024).

## 3 Methodology

Based on these recent findings, it is our goal to keep investigating the research data submitted to and published at the DELFI conference. We further intend to analyze what researchers perceive as their data, and what they publish as such. The research questions are as follows:

> RQ1: *Which types of research data are (not) recognized as such by DELFI 2024 authors?*
>
> RQ2: *What research data was published at the DELFI 2024 conference?*
>
> RQ3: *Where is research data from DELFI 2024 papers published?*

To answer our research questions, we utilize two different data sources. RQ1 is addressed via a two-step process. First, we analyze the submission data from EasyChair for the DELFI 2024 conference. Thus, we investigate the authors' intentions to publish their research data for all 125 submissions (regardless of their acceptance or rejection). The answer options were as follows: (a) "no data to publish", (b) "I will publish the data", (c) "I don't want to", and (d) "I already published the data". Second, we analyzed all submissions tagged with "no data to publish" and compared the authors' views and/or intentions with our assessment of their submissions. That is, we read the submissions and extracted the research methods. In cases where the authors had explicitly described their methods, we also extracted a description of the research methods and data. When the description was abstract (e.g., only mentioning "interviews" or a "prototype") the research paradigm and type of data (e.g., "qualitative" and "software") were added. In addition, the following criteria for the identification of research data were applied. The data had to be explicitly linked in the submission. We only recognized data that authors had gathered or developed themselves, meaning instruments or large task pools developed by others were not considered. All submissions that claimed to have no data were reviewed by both authors, and difficult cases were discussed until reaching a consensus. For the classification, we further consulted existing literature on research data types, e.g., (Biernacka, Mulligan, et al. 2023; Kiesler, Röpke, et al. 2024; Kiesler, Impagliazzo, et al. 2024).

For RQ2 and RQ3, we use the DELFI 2024 conference proceedings as a basis (Schulz and Kiesler 2024). We analyzed all 55 contributions from all categories (11 long, 17 short, 11 practice and 1 position paper, 11 demos, 4 posters).

The analysis focuses on the following elements: For RQ2, type of research data (e.g., qualitative, quantitative, or software in analogy to RQ1); and for RQ3 where the publication is stored (e.g., online repository, project website, etc.).

## 4 Results

### 4.1 Recognition of Research Data – RQ 1

When authors submitted their papers for peer-review (see Fig. 1), 9 authors declared they already published their research data. Authors of 38 submissions indicated they wanted to publish their research data. Also 22 author teams had no intention to publish their research data. Of all 125 submissions, 56 authors declared not to have research data to publish (45%).

However, in 35 submissions (out of 56 submission tagged as "no data to publish") we identified research data. They were described as being part of (mostly empirical) studies. The type of research

|                              | full | short | position | practice | demo | poster | sum | percent |
|------------------------------|------|-------|----------|----------|------|--------|-----|---------|
| No data to publish           | 7    | 15    | 5        | 13       | 8    | 8      | 56  | 45      |
| No intention to publish      | 9    | 5     | 0        | 5        | 1    | 2      | 22  | 18      |
| Intention to publish in the future | 14 | 14 | 0      | 4        | 3    | 3      | 38  | 30      |
| Already published the data   | 2    | 3     | 0        | 1        | 3    | 0      | 9   | 7       |

Created with Datawrapper

Figure 1: Authors' Perspective on Publication of Their Research Data

data that seemed to be largely overseen was *qualitative data* ($N$ = 20). Similarly, *software* was often ($N$ = 19) not categorized as research data. *Quantitative data* was also overlooked ($N$ = 8). In the next paragraphs we will dive deeper into the concrete data kinds of overseen data.

- **Qualitative data:** Examples for some common research data found are interview studies, text analyses, or qualitative parts of surveys. Many more kinds of qualitative research data was found, e. g. teaching concepts, teaching materials (e. g. in Moodle courses), lecture notes, exercises, or ChatGPT promotions.

- **Software:** Many research data in this category were developed or adapted LLM or prototype for VR/AR applications. In addition, a large variety of developed software tools were included in the submitted articles.

- **Quantitative data:** For this category, we found in the submitted articles e. g. usability studies, assignment tagging, surveys, click counts, which are quantitative in nature.

## 4.2 Research Data Published at the DELFI 2024 Conference – RQ2 and RQ3

**RQ2: What research data?** – The analysis of the papers accepted at the DELFI 2024 conference led to the identification of 17 submissions with openly available research data. Among them are 5 full papers, 6 short and 3 practice papers, and 3 demos. None of the four published posters had published their data. Regarding the type of data, we identified data from the qualitative and quantitative paradigms, but also software and code (e.g., prototypes, analysis scripts, etc.). In some publications, we found data from multiple research paradigms, for example, interview guides and transcripts, mock-ups, code, and survey data (instruments and responses). For this reason, the numbers presented in Table 1 are higher than 17; they add up to 20. Overall, software was the most frequent type of research data, with 11 publications overall, followed by 6 qualitative, and 3 quantitative datasets.

**RQ3: Where published?** – The research data within the DELFI 2024 proceedings were published at GitHub/GitLab ($N$ = 9), OSF ($N$ = 6), Zenodo ($N$ = 1), OneDrive Cloud ($N$ = 1), or in the Appendix of the paper ($N$ = 1). Almost all papers used one repository/space for publishing their data – except for two papers where the authors provided two links to their research data (one to OSF and one to GitLab). The source of one dataset could not be identified, as the hyperlink in the paper had been shortened and expired. For two other datasets, we noted that links were invalid, or that projects had been moved to other folders.

|             | full | short | practice | demo |
|-------------|------|-------|----------|------|
| Qualitative | 2    | 3     | 1        | 0    |
| Software    | 3    | 3     | 2        | 3    |
| Quantitative| 2    | 1     | 0        | 0    |

Table 1: Types of data publications at the DELFI 2024 conference.

## 5 Discussion

The evaluation of EasyChair's information along with the analysis of published data revealed insights into the DELFI community's perspective on research data. We will discuss interesting observations, and relate them to prior work. For example, the results of RQ 1 uncovered potential research data, which were not specified by the authors as such. It is a paradox that within the EdTec Community, *software* is not recognized as research data (cf. Kiesler and Schiffner 2022). It is possible though that some authors had already published their prototypes, but lacked awareness of their importance within the epistemological process. Another aspect may be that publishing research data is perceived as too complex or demanding; guidelines and support for authors may be unknown (cf. Kiesler, Röpke, et al. 2024). The latter particularly applies to PhDs. General guidelines from the research data management community do exist (Biernacka, Bierwirth, et al. 2020). However, we might need to increase our efforts to link the communities (cf. Biernacka and Schulz 2022).

Another observation was that *qualitative research data* often remained unpublished. The reasons are likely manifold. For example, authors may lack awareness that their data (e.g., course material, teaching concepts) can be valuable for others. Other concerns may be proper anonymization and alignment with ethics board approval. However, the qualitative research process is rich (cf. Brown and Guzdial 2024; Keuning 2024) in information and details should be published to support sustainable interaction with research resources.

It is worth discussing the short viability of some research data publications, which was particularly due to broken hyperlinks (instead of DOIs), or moved project folders. Considering the timing of submissions for the proceedings (06/2024), and the time of this analysis (04/2025) the brevity is deeply concerning. It raises serious questions regarding accessibility and FAIR principles (Wilkinson et al. 2016), and confirms previous findings from Melkamu and Striewe (2023). To date, the DELFI conference does not have a comprehensive policy in place to motivate, incentivize, and guide authors towards publishing their research data.

**Limitations:** The following limitations of this study should be noted. For example, the analysis was only conducted for DELFI 2024 submissions and publications. To gain deeper insights into the data and publishing behaviour of the DELFI community, more data needs to be gathered and analyzed – preferably over multiple years and conference iterations. In addition, we note the multifaceted reasons and barriers for researchers leading to the decision to not publish their data (see, e.g., Kiesler, Röpke, et al. 2024). Thus, authors who did not publish research data do not necessarily lack awareness towards them. There may have been other reasons why the authors did not reveal their intentions during the submission process through EasyChair.

## 6 Conclusions and Future Work

The analysis of the DELFI 2024 conference submissions and proceedings w.r.t. the publication of research data revealed the community's data practices. Few authors (7 %) had published their data upon submitting to DELFI, 19 % had no intention to do so, and 30 % wanted to publish them in the future. 45 % of submissions ($N = 56$) were labeled as not having data to share. However, we identified research data for 35 of them - above all qualitative data ($N = 20$), and software ($N = 19$), but also quantitative data ($N = 8$). The 17 accepted submissions that had their data published comprised mostly software ($N = 11$), but also qualitative ($N = 6$) and quantitative data ($N = 3$). It was available via GitHub/GitLab ($N = 9$), OSF ($N = 6$), Zenodo ($N = 1$), OneDrive Cloud ($N = 1$), and an Appendix ($N = 1$). Unfortunately, three datasets could not be accessed. These findings have several implications for the DELFI community, as they emphasize the urgent need for guidelines and recommendations regarding the publication of research data. We need to align and promote our understanding of research data (cf. Striewe 2024) and how to publish them. This encompasses recognition of existing guidelines (e.g., Biernacka, Bierwirth, et al. 2020), and training future EdTec researchers. This training should begin during study programs (Schulz and Jacob 2025; Petersen et al. 2023). To monitor this process, we recommend follow-up studies focusing on the submission and publication of research data at the DELFI, and other EdTec conferences.